\documentclass[a4paper,11pt]{article}
\usepackage{pos}

\title{Systematic uncertainty of offshell corrections and higher-twist contribution in DIS at large x}

\ShortTitle{Systematic uncertainties in DIS at large x}

\author*[a,b]{Matteo Cerutti}
\emailAdd{mcerutti@jlab.org}
\affiliation[a]{Hampton University, \\ Hampton, Virginia 23668, USA}
\affiliation[b]{Jefferson Lab, \\ Newport News, Virginia 23606, USA}

\author[b,c]{Alberto Accardi}
\emailAdd{accardi@jlab.org}
\affiliation[c]{Christopher Newport University \\ Newport News, Virginia 23606, USA}

\author[d]{Ishara P. Fernando}
\emailAdd{ishara@virginia.edu}
\affiliation[d]{University of Virginia, \\ Charlottesville, Virginia 22904, USA}

\author[e]{Shujie Li}
\emailAdd{shujieli@lbl.gov}
\affiliation[e]{Lawrence Berkeley National Laboratory, \\ Berkeley, California 94720, USA\\ \ }

\abstract{We study the systematic uncertainty and biases introduced by theoretical assumptions needed to include large-$x$ DIS data in a global QCD analysis. Working in the CTEQ-JLab framework, we focus on different implementations of higher-twist corrections to the nucleon structure functions and of offshell PDF deformations in deuteron targets and discuss how their interplay impacts the extraction of the $d$-quark PDF and the calculation of the neutron structure function at large $x$.}

\FullConference{31st International Workshop on Deep Inelastic Scattering (DIS2024)\\
 8–12 April 2024\\
Grenoble, France\\}

\renewcommand{\logo}{\relax}

\renewcommand{\hookAfterAbstract}{%
\par\bigskip
\textsc{JLab preprint number}: JLAB-THY-24-4100
}

\begin{document}
\maketitle

\section{Introduction}
Knowledge of parton distributions functions (PDFs) at large longitudinal momentum fraction $x$ is crucial to characterize the effects of color confinement and nonperturbative interactions on the proton's structure and to search for beyond the Standard Model interactions in large mass and forward particle production at the Large Hadron Collider \cite{Accardi:2015lcq}. 

In a global QCD analysis, such as from the CTEQ-JLab (CJ) collaboration \cite{Accardi:2016qay,Accardi:2023gyr}, one can combine diverse experimental data to constrain the large-$x$ behavior of the PDFs flavor by flavor. For the $u$-quark, information can be gathered from the large amount of Deep-Inelastic Scattering (DIS) data on proton targets,
from  fixed-target Drell--Yan data, and on jet production in hadron-hadron collisions.
For the $d$-quark, strong constraints come from precision data on large rapidity $W$-boson asymmetries in proton-antiproton collisions, and from abundant inclusive DIS data on \textit{deuteron} targets, as well as from pioneering proton tagged deuteron DIS at JLab.  
An accurate description of deuteron DIS data, which are predominantly taken at small energy scale $Q^2$, requires one to correct the theoretical calculations for the nuclear dynamics of the target nucleons -- including nuclear binding, Fermi motion, and nucleon offshell deformation -- and for $1/Q^2$ power-suppressed effects. The latter include $O(M^2/Q^2)$ kinematic target mass corrections (TMCs), $O(\Lambda_\text{QCD}^2/Q^2)$ dynamical effects such as multiparton correlations, and any other ``residual'' $1/Q^2$ corrections, say, from higher-order QCD diagrams \cite{Blumlein:2008kz} or unaccounted for mass corrections. While TMCs can be calculated, the remaining terms are generally fitted to data and collectively named ``higher-twist (HT) corrections''. Likewise, parametrized offshell deformations in nuclear targets can be constrained by leveraging, through a global fit of the underlying PDFs, $d$-quark sensitive data on free proton targets such as the above mentioned $W$ rapidity asymmetry. 

In this contribution, we focus on the interplay of the offshell and higher-twist corrections. We show that the implementation choices for HT corrections may introduce a bias in the calculation of the large-$x$ behavior of the DIS structure functions, and argue that this bias can in turn be partially compensated by the fitted offshell effects. Working consistently in the CJ22 framework \cite{Accardi:2023gyr}, we demonstrate the presence of such bias in an actual PDF fit, identify safe implementations of the HT corrections, and discuss the related systematic uncertainty. We also briefly comment on similarities with studies of offshell nucleon deformations performed by Alekhin, Kulagin and Petti (AKP) \cite{Alekhin:2022tip,Alekhin:2022uwc} and by the JAM collaboration \cite{Cocuzza:2021rfn}.

\section{Formalism}
\label{s:Formalism}

The deuteron $F_{2D}$ structure function can be written in the nuclear impulse approximation as
\begin{equation}
F_{2D} (x, Q^2) = \int d y d p_T^2 f_{N/D}(y, p_T^2; \gamma) F_{2N} \bigg (\frac{x}{y}, Q^2, p^2 \bigg ).
\label{e:SIA}
\end{equation}
Here $x=2 Q^2 / P_D\cdot q$ is the \textit{per-nucleon} Bjorken invariant, with $q$ and $P_D$ the photon and deuteron momenta, respectively. $F_{2N}$ is the structure function of an offshell nucleon $N$, $x/y$ its Bjorken invariant, and $\gamma = \sqrt{1 + 4 \frac{x^2}{y^2} \frac{p^2}{Q^2}}$. The bound nucleon squared 4-momentum, $p^2$, is in general smaller than the squared on-shell mass $M_N^2$. 
The smearing function $f_{N/D}$ is built out of the nucleon wave function in the deuteron. 

Since the deuteron is weakly bound, we can Taylor-expand $F_{2D}$ around its on-shell limit~\cite{Kulagin:1994cj,Kulagin:1994fz}. This can be done at the parton level or at the structure function level, namely,
\begin{align}
    q(x,Q^2,p^2) &= q^{\text{free}}(x,Q^2) \big ( 1 + \frac{p^2 - m_N^2}{m_N^2} \delta f (x) \big ) \,
    \label{e:off_pdf}
    \\
     F_{2N}(x,Q^2,p^2) &= F_{2N}^{\text{free}}(x,Q^2) \big ( 1 + \frac{p^2 - m_N^2}{m_N^2} \delta F (x) \big ) \,
      \label{e:off_sf}
\end{align}
where $q^{\text{free}}$ and $F_{2N}^\text{free}$ are free-nucleon PDFs and structure functions, and the $\delta f$, $\delta F$ ``offshell function'' parametrize the partons or nucleon deformation when bound in a nucleus. 
With an isospin-symmetric target such as the deuteron, the offshell functions have to be flavor independent; with the addition of data on $A=3$ nuclei, the isospin dependence can also be studied \cite{Cocuzza:2021rfn,Alekhin:2022uwc}. As in CJ22
we implement the offshell corrections at partonic level, but parametrize $\delta f$ with a generic polynomial of second degree to reduce parametrization bias. A polynomial was also used by AKP for the structure function level $\delta F$~\cite{Alekhin:2017fpf}.  
We also note that, even in the presetn flavor-independent implementation, $\delta f$ and $\delta F$ are different beyond LO and should be compared with care. 

HT corrections are usually implemented in phenomenological QCD studies in terms of \textit{mutiplicative} or \textit{additive} $x$-dependent functions:
\begin{align}
    F_2^\text{mult}(x,Q^2) &= F_2^\text{TMC}(x,Q^2) \bigg ( 1 + \frac{C(x)}{Q^2} \bigg ) \ , 
    \label{e:ht_mult}
    \\
      F_2^\text{add}(x,Q^2) &= F_2^\text{TMC}(x,Q^2) + \frac{H(x)}{Q^2} \ , 
    \label{e:ht_add}
\end{align}
with TMCs applied to the LT structure function on the right hand side.
The only difference in the two implementations is in the assumed $Q^2$ evolution of the HT function, since the multiplicative implementation \eqref{e:ht_mult} can be rewritten in additive terms 
\begin{equation}
    F_2^\text{mult}(x,Q^2) = F_2^\text{TMC}(x,Q^2) + \frac{\tilde{H}(x,Q^2)}{Q^2} \ ,
    \label{e:ht_tilde}
\end{equation}
with a $Q^2$ evolution inherited by the $F_2$ structure function, namely, $\tilde{H}(x,Q^2) = F_2(x,Q^2) C(x)$. Due to the relative scarcity of large-$x$ data, the HT terms are typically approximated as isospin symmetric (see, \textit{e.g.}, Refs.~\cite{Alekhin:2017fpf,Accardi:2016qay}) with older studies indicating indeed little sensitivity in the data to isospin symmetry breaking~\cite{Alekhin:2003qq}. Nevertheless, as shown in this contribution, some discriminating power exists when including JLab 6 GeV data. Isospin-breaking HT terms have also been considered in a recent fit by the JAM collaboration \cite{Cocuzza:2021rfn}.

The HT implementation choices may give rise to systematic uncertainty in the QCD analysis, especially at large $x$ where PDFs steeply fall to 0.
We can look, for example, at the neutron-to-proton ratio of the $F_2$ structure functions, which we denote with $n/p$, whose limiting behavior as $x \to 1$ is sensitive to confinement effects but cannot be directly measured. This ratio may instead be inferred from proton and deuteron target data after removing nuclear corrections: for example, by calculating it in pQCD with the PDFs obtained in the global analysis as we do here, or, similarly, in a data-driven analysis such as in Ref.~\cite{Li:2023yda}. In a global fit, however, both HT and offshell corrections affect the determination of the deuteron structure function, and biases in either one can be compensated by the fitted parameters of the other, potentially leading to very different determinations of the $n/p$ ratio.

If we neglect target-mass corrections, the $n/p$ ratio at LO and leading twist reads
\begin{equation}
    \frac{n}{p} \stackrel{x \rightarrow 1}{ \xrightarrow{\hspace{1cm}} } \frac{4d + u}{4u + d} \simeq \frac{1}{4} 
    \label{e:nop_std}
\end{equation}
using $d/u \to 0$.
With isospin-symmetric, \textit{multiplicative} HT functions ($C(x) \equiv C_p(x) = C_n(x))$, the correction simply cancels in the $n/p$ ratio and one obtains the same limit as in Eq.~\eqref{e:nop_std}. For isospin-symmetric \textit{additive} HT corrections ($H(x) \equiv H_p(x) = H_n(x)$) we obtain, instead,
\begin{equation}
    \frac{n}{p} \stackrel{x \rightarrow 1}{ \xrightarrow{\hspace{1cm}} } \frac{u + H/Q^2}{4u + H/Q^2} \simeq \frac{1}{4} + 3 \frac{H}{16uQ^2} \ ,
    \label{e:nop_Aiso}
\end{equation}
where we used $d/u \to 0$ again, and in the last step we performed a Taylor expansion in the small $H / (uQ^2)$ term. The additive HT produces a larger tail than the multiplicative HT purely due to the phenomenological implementation choice, potentially overestimating the $n/p$ ratio. 
Since the neutron $F_{2n}$ structure function only enters in deuteron measurements, this tail can be effectively compensated in the fit by a positive offshell $\delta f$ (or $\delta F$) function. Conversely, the multiplicative implementation may lead to an underestimate of $F_{2n}$ that can be compensated by a negative offshell deformation, with a $\chi^2$ values comparable to the additive case. 

This implementation bias can be removed by considering isospin breaking for the HT terms. 
Indeed, consider that one can go from the additive to the multiplicative representation by the $H_{p,n} \to \tilde H_{p,n}$ substitution, we obtain 
\begin{equation}
    \frac{n}{p} \stackrel{x \rightarrow 1}{ \xrightarrow{\hspace{1cm}} } \frac{u + \tilde{H}_n/Q^2}{4u + \tilde{H}_p/Q^2} \simeq \frac{1}{4} + \frac{4\tilde{H}_n - \tilde{H}_p}{16uQ^2} \simeq \frac{1}{4} + \frac{\tilde{H}_p}{16uQ^2} \ ,
    \label{e:nop_noiso}
\end{equation}
in either representation. In the last step, we estimated $\tilde H_n \approx \frac12 \tilde H_p$ according to Ref.~\cite{Alekhin:2003qq}. As a result, the bias in the isospin-symmetric implementation of Eqs.~\eqref{e:nop_std}-\eqref{e:nop_Aiso} is removed, and the $n/p$ tail in Eq.~\eqref{e:nop_noiso} is closer to the multiplicative estimate. With isospin-breaking HT corrections, we can also expect that the fitted offshell function will not dependent on the HT implementation. 


\section{Results}

In order to corroborate our theoretical expectations, we have implemented the HT scenarios just discussed in the CJ22 global analysis framework ~\cite{Accardi:2023gyr}, but using a second order polynomial for added flexibility in the parametrization of the parton-level $\delta f$ offshell function.

\begin{figure}[tbh]
\centering
\includegraphics[width=0.97\textwidth]{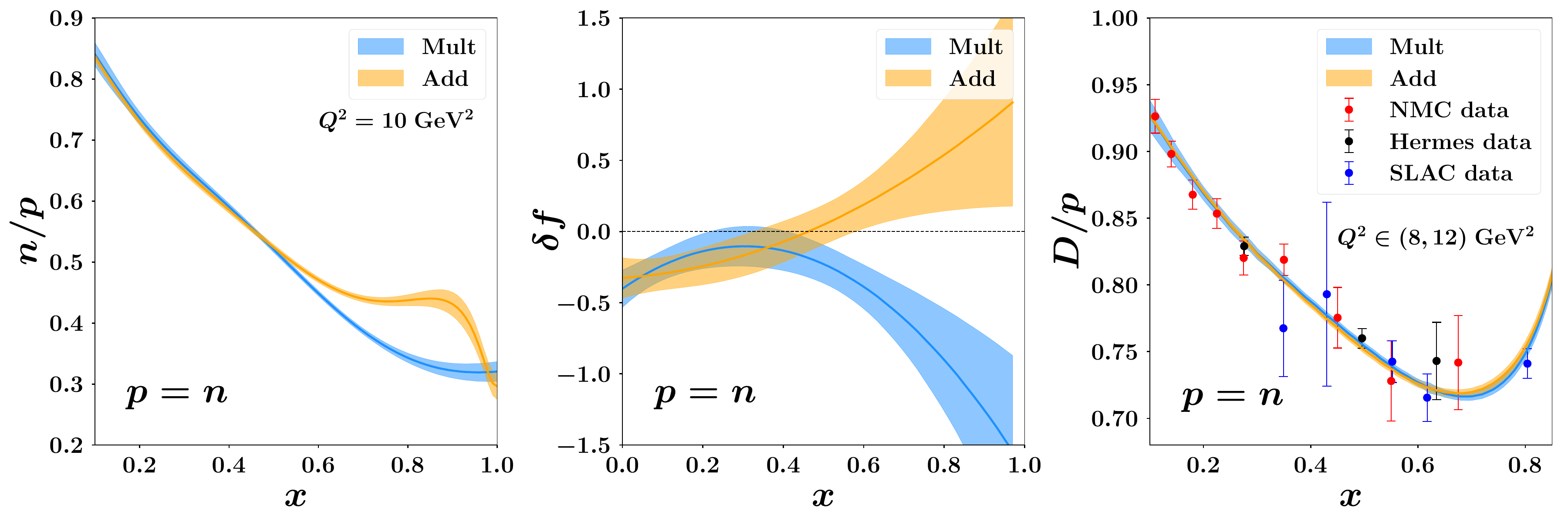}
\vskip-0.2cm
\caption{
Comparison of the results of the CJ analyses when implementing isospin-symmetric ($p=n$) additive (orange band) or multiplicative (violet band) HT corrections. \textit{Left panel}: $n/p$ ratio of $F_2$ structure functions at $Q^2=10$ GeV$^2$. \textit{Central panel:} offshell function. \textit{Right panel}: $D/p$ ratio of $F_2$ structure functions at $Q^2=10$ GeV$^2$ compared to a selection of experimental data.
Bands represent 1-sigma uncertainties.}
\label{f:HT_iso}
\end{figure}

\begin{figure}[t]
\centering
\includegraphics[width=0.97\textwidth]{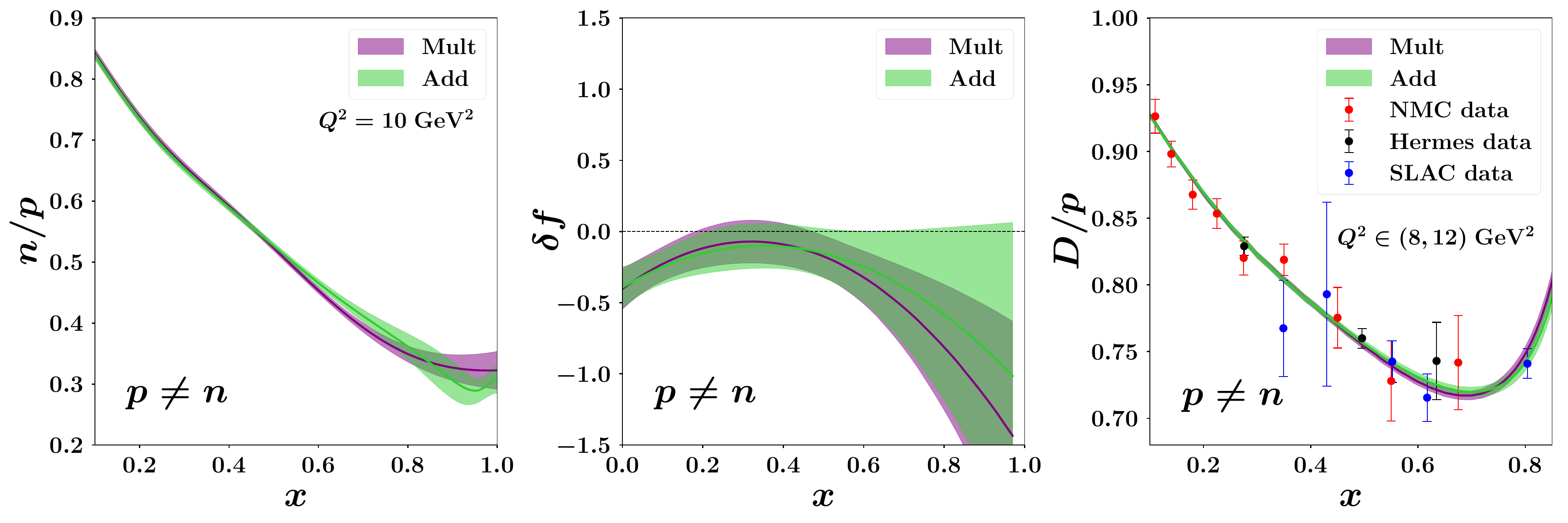}
\vskip-0.2cm
\caption{
Comparison of the isospin-breaking ($p\neq n$) additive (green band) or multiplicative (purple band) implementation of the HT corrections in the CJ global analysis. Curves and bands as in Fig.~\ref{f:HT_iso}.}
\vskip-0.2cm
\label{f:HT_noiso}
\end{figure}

In Fig.~\ref{f:HT_iso}, we show fits performed with isospin-symmetric HT corrections.
We observe a pronounced difference between the extracted $n/p$ ratios in the left panel, with a substantially higher tail for the additive implementation up in the $x < 0.85$ region covered by the data, after which the HT parametrization actually forces the ratio to go 0. At the same time, the description of the $D/p$ experimental data is rather stable, see the right panel. 
In Fig.~\ref{f:HT_noiso}, we display the results for the isospin-breaking HT implementation. The $n/p$ ratio extracted with additive and multiplicative HT corrections now agree with each other and present a tail which is in between those in Fig.~\ref{f:HT_iso} (and closer to the multiplicative one, as expected).
No compensation by the offshell function is then needed to properly describe the experimental data on $D/p$, and the extracted $\delta f$ functions are compatible with each other.

Comparison of Figs.~\ref{f:HT_iso} and \ref{f:HT_noiso} demonstrates the implementation bias discussed in the previous section and summarized in Eqs.~\eqref{e:nop_std}-\eqref{e:nop_noiso}: the isospin-symmetric implementation of additive HT correction artificially increases the size of the $n/p$ tail at large $x$, while the multiplicative implementation suppresses it. This artificial behavior can, however, be compensated by a large and positive offshell correction at large $x$ in the additive case\footnote{A large and positive offshell function $\delta f$ generates a smaller structure function, see Eq.~\eqref{e:off_sf}.}, and by a negative one in the multiplicative case. In the isospin-breaking scenario the bias is removed: the extracted $n/p$ ratios and $\delta f$ offshell deformations become stable against the choice of HT representation, and the fitted $H_{p,n}$ and $\delta$ functions can be interpreted with confidence. The small remaining differences can then be used to estimate the (small) HT implementation systematic uncertainty. 

In the central panel of Fig.~\ref{f:HT_noiso}, we observe that the partonic offshell functions we have extracted are compatible with 0, except at small $x$. Does this mean that the quark densities are not modified by nucleon offshell effects? Not really, because a non-zero offshell deformation of the $u$ quark maybe be cancelled by an opposite deformation of the $d$ quark, as shown in a recent JAM analysis \cite{Cocuzza:2021rfn}. In order to perform a similar flavor decomposition of the $\delta f$ function, we would also need to include in the fit the \textsc{MARATHON} experimental DIS data on $^3H$ and $^3He$ targets from Jefferson Lab~\cite{JeffersonLabHallATritium:2021usd}. 

\begin{figure}[tbh]
\centering
\includegraphics[width=0.8\textwidth]{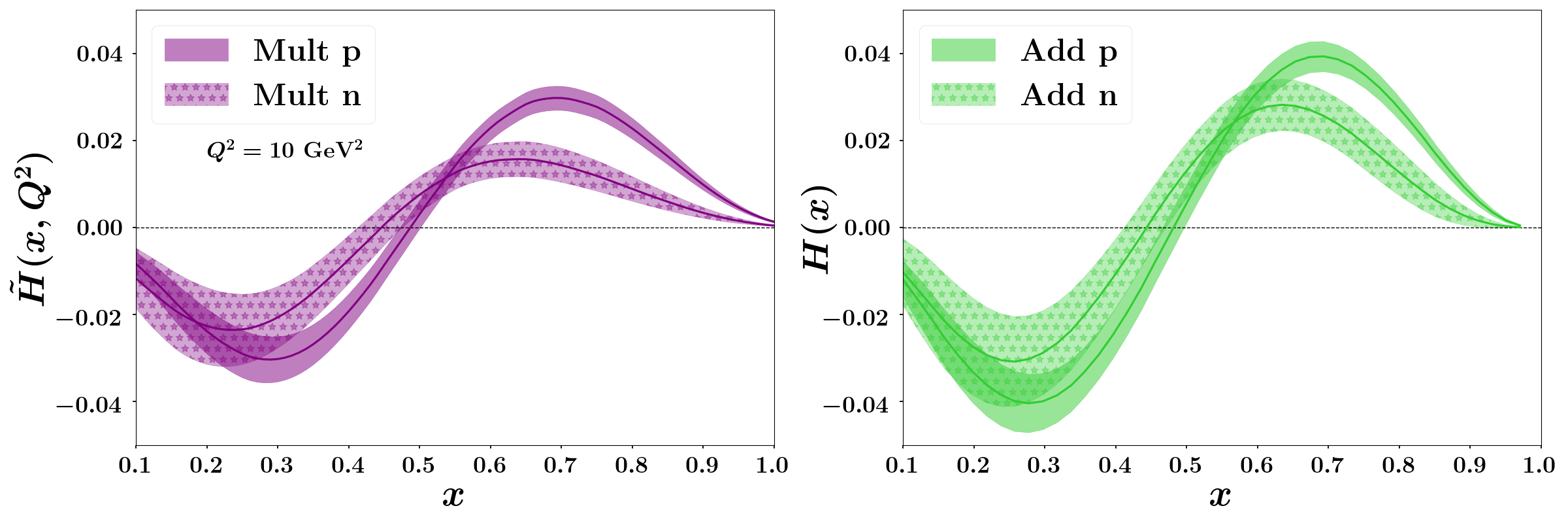}
\caption{
Comparison of proton (p) and neutron (n) HT extractions in isospin-breaking fits. \textit{Left}: multiplicative HT implementation. \textit{Right}: additive HT implementation. The bands represent 1 sigma uncertainties.}
\label{f:fitted_HT}
\end{figure}

In Fig.~\ref{f:fitted_HT}, we show the extracted HT terms in the isospin-breaking fit and compare the additive (right panel) and multiplicative (left panel) implementations.
We note that beyond $x =0.5$ the proton's HT is larger than and distinct from the neutron's HT, showing the discriminating power of the DIS data set included in our analysis. Note also that $H_n(x) \simeq 1/2 H_p(x)$ at large $x$, as assumed in the derivation of Eq.~\eqref{e:nop_noiso}.

A study of the additive and multiplicative HT implementations has also been performed by AKP in Refs.~\cite{Alekhin:2022tip,Alekhin:2022uwc}, but limited to the isospin-symmetric case. At variance with our results, AKP find no significant impact of the choice HT implementation: in both cases, their $n/p$ ratio and structure function-level $\delta F$ have a shape similar to our additive fits shown in orange in Fig.~\ref{f:HT_iso}.
While it is difficult to analyze the intricate interplay of the many elements and implementation choices in a global QCD analysis by other groups, we notice that they do obtain a statistically significant variation in the large-$x$ tail of the $d/u$ PDF ratio when implementing additive or multiplicative higher twists (see Figure 4 of Ref.~\cite{Alekhin:2022uwc}). This happens at $x>0.3$, where the statistical power of the decay lepton asymmetry in $W$ boson production wanes \cite{Accardi:2016qay}, leaving room for the fit to compensate the HT bias with a deformation of the $d$ quark PDF. In comparison, CJ22 also leverages data on kinematically reconstructed $W$ boson asymmetry, which provide strong constraints on the $d$ quark up to $x \sim 0.7$, therefore shifting the brunt of the bias compensation to the offshell function.


\section{Conclusions}

The observed discrepancies in the offshell deformations fitted by AKP, JAM and CJ can likely be traced back to the differences in the phenomenological implementation of nuclear effects and higher-twist corrections in the different computational frameworks,as well as, possibly, the choice of data.
These discrepancies also highlight the necessity of performing a detailed study of the systematic effects and possible biases induced by the many phenomenological choices one has to make when performing a global QCD analysis, that are not necessarily limited to those addressed in this contribution. New experimental data from Jefferson Lab, in particular of tagged DIS cross sections binned in the spectator four-momentum squared, will be  helpful in improving the robustness of global QCD analysis frameworks and in correctly characterizing the off-shell deformation of bound nucleons. 


\section*{Acknowledgments}

\noindent
This work was supported in part by the  U.S. Department of Energy (DOE) contract DE-AC05-06OR23177, under which Jefferson Science Associates LLC manages and operates Jefferson Lab, and by DOE contract DE-SC0008791.

\end{document}